\documentclass[aps,pra,amsfonts,amssymb,epsfig,twocolumn,groupedaddress,showpacs]{revtex4}
\usepackage{graphics}
\usepackage{epsfig}
\usepackage{amsmath}
\newcommand{\Tr}{\mbox{Tr}}

\begin{document}
\title{Degradability of Bosonic Gaussian channels}
\author{Filippo Caruso and Vittorio Giovannetti}
\affiliation {NEST CNR-INFM $\&$ Scuola Normale Superiore, Piazza
dei Cavalieri 7, I-56126 Pisa, Italy}
\date{\today}

\begin{abstract}
The notion of weak-degradability of quantum channels is introduced
by generalizing the degradability definition given by Devetak and
Shor. Exploiting the unitary equivalence with
beam-splitter/amplifier channels we then prove that a large class
of one-mode Bosonic Gaussian channels are either weakly degradable
or anti-degradable. In the latter case this implies that their
quantum capacity $Q$ is null. In the former case instead, this
allows us to establish the additivity of the coherent information
 for those maps which admit unitary
representation with single-mode pure environment.

\end{abstract}
\pacs{03.67.Hk, 42.50.Dv} \maketitle

\section{Introduction}
Bosonic Gaussian channels provide a realistic noise model for
transmission lines  which employ photons as information carriers
including optical fibers, wave guides, and free-space e.m.
communication. They account for all processes where the
transmitted signals undergo loss, amplification, and/or squeezing
transformations~\cite{HW,HOLEVOBOOK}. The characterization of
these transmission lines is relevant not only  from a
technological point of view but also from the point of view of
quantum information theory where they pose  some important open
problems (see Ref.~\cite{REV1} for a review). For instance, in the
context of average input photon number constraint, it is believed
that the optimal (classical or quantum) communication
rates~\cite{SHOR} of such  channels should be achieved
 by encoding messages into Gaussian
input states~\cite{HW,HALL,ENTROPY,SHAPIRO}. However, apart from
the noiseless case~\cite{CAVES}, the only nontrivial map for which
such conjecture has been proved is the purely lossy channel where
the information carrying photons couple through beam splitters
with an external vacuum state (see  Ref.~\cite{LOSS} for the
classical capacity~\cite{HSW} case and Refs.~\cite{BROAD,HW} for
the entangled assisted capacities). Finally Bosonic Gaussian
channels are generally believed~\cite{HW} to provide a natural
example of maps with additive properties~\cite{SHORADD} (e.g., it
has been conjectured that their maximum Holevo information and
minimum R\'{e}nyi entropies should be additive) although only
preliminary results have been obtained~\cite{GL,SEW,LOSS} so far.
In this paper we discuss a  property which has some relevant
implications in the analysis of the quantum capacity
$Q$~\cite{SETH} of a large class of one-mode  Bosonic Gaussian
channels. Indeed we show that these maps are either {\em weakly
degradable} or {\em anti-degradable}. The notions of
weak-degradability and anti-degradability  of a channel are
introduced here as  a generalization of the degradability property
defined by Devetak  and Shor~\cite{DEVSHOR}. On one hand, using an
argument of Refs.~\cite{ERASURE,BROAD,SARO}, one can show that
anti-degradable channels have null quantum capacity. On the other
hand, it is known that channels which are degradable in the sense
of Ref.~\cite{DEVSHOR} possess  additive coherent information
which allows one to express their quantum capacity, $Q$, in terms
of a single-letter formula. The latter is much easier to handle
than its regularized version and, in some cases, allows for a
complete characterization of the $Q$ (e.g., see the dephasing
channel in Ref.~\cite{DEVSHOR} and amplitude damping channel in
Ref.~\cite{SARO}). Weakly degradable maps do not necessarily share
the above property (at present, however, we do not have a
counter-example of this fact). Still, as will be clear in the
following, weak-degradability provides a necessary condition for
degradability and for some of the maps analyzed here the two
notions coincide. In what follows, we first present the formal
definitions of weak-degradability and anti-degradability (see
Sec.~\ref{SEC:WDADM}). Then in Sec.~\ref{SEC:ONEMODE}
 we  introduce one-mode Bosonic Gaussian channels
and in Sec.~\ref{sec:bsandamp} we analyze their degradability
properties.

\begin{figure}[t]
\begin{center}
\epsfxsize=1 \hsize\leavevmode\epsffile{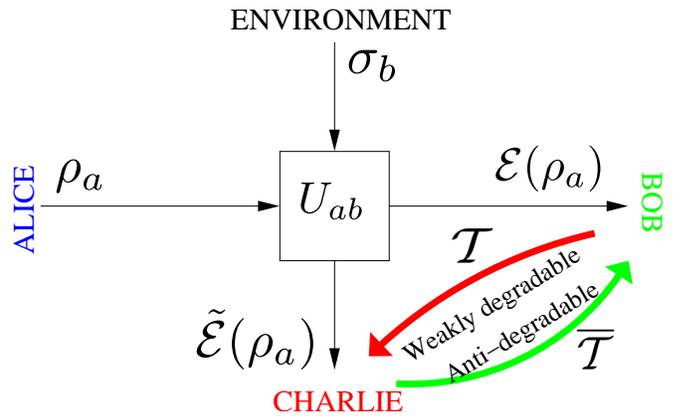}
\end{center}
\caption{Weakly degradable vs. anti-degradable channels. A channel
$\cal E$ is weakly degradable if there exists a CPT map $\cal T$
which, for all input $\rho_a$, allows Bob to recover the
environment output $\tilde{\cal E}[U_{ab},\sigma_b](\rho_a)$ from
${\cal E}(\rho_a)$ as in Eq.~(\ref{deg}). A channel $\cal E$ is
anti-degradable if, instead,
 ${\cal E}(\rho_a)$ can be recovered from  $\tilde{\cal E}[U_{ab},\sigma_b]
(\rho_a)$ via a CPT transformation $\overline{\cal T}$ as in
Eq.~(\ref{antideg}). Weak-degradability reduces to the
degradability condition of Ref.~\cite{DEVSHOR} if, in
Eq.~(\ref{unit}), the environment $\sigma_b$ is pure.}
\label{f:figu1}
\end{figure}

\section{Weakly Degradable and Anti-Degradable maps}\label{SEC:WDADM}
Consider  a quantum channel described by the completely positive,
trace preserving (CPT) map ${\cal E}$. It can always be
represented as a unitary interaction, $U_{ab}$, between the state,
$\rho_a \in {\cal D}({\cal H}_a)$, of  the channel information
carrier and an external environment initially prepared in a
(generally mixed) state, $\sigma_b \in {\cal D}({\cal H}_b)$, i.e
\begin{eqnarray}
{\cal E}(\rho_a) = \Tr_b[ U_{ab} (\rho_a\otimes \sigma_b)
U_{ab}^\dag] \label{unit} \;,
\end{eqnarray}
where $\Tr_b[ ...]$ is the partial trace over the environment and
${\cal D}({\cal H}_{a,b})$ are the sets of the density matrices of
the system $a$ and $b$, respectively. For $\sigma_b$ pure
Eq.~(\ref{unit}) is equivalent to the Stinespring
representation~\cite{STINE} of ${\cal E}$ and its properties are
uniquely determined up to an isometry (i.e., they do not depend
upon the possible choices of  $U_{ab}$ and $\sigma_b =
|\psi\rangle_b\langle\psi|$). For $\sigma_b$ mixed such unicity is
generally lost. For this reason in the following  we will write
${\cal E}[U_{ab},\sigma_b]$ instead of ${\cal E}$ to make explicit
which  representation~(\ref{unit}) is under consideration. We  now
introduce the CPT map $\tilde{\cal E}[U_{ab},\sigma_b]$ which
takes $\rho_a$ into the density matrix which describes the
environment after the interaction with the carrier, i.e.,
\begin{eqnarray}
 \tilde{\cal E}[U_{ab},\sigma_b] (\rho_a) \equiv
\Tr_a[ U_{ab} (\rho_a\otimes \sigma_b) U_{ab}^\dag]
\label{unitKconjugate}
\;,
\end{eqnarray}
where $\Tr_a[ ...]$ is the partial trace over the carrier (see
also Fig.~\ref{f:figu1}). Since we are not assuming $\sigma_b$ to
be pure, Eq.~(\ref{unitKconjugate}) differs from the standard
definition of conjugate map given in
Refs.~\cite{DEVSHOR,KING,HOLEVONEW}. For this reason we name the
map $\tilde{\cal E}[U_{ab},\sigma_b]$ as weakly complementary or
weakly conjugate of ${\cal E}$ with respect to the
representation~(\ref{unit}).

A channel $\cal E$  is said to be weakly degradable with respect
to the representation~(\ref{unit}) if given an unknown carrier
input state $\rho_a$, the receiver of the messages (Bob) can
reconstruct the perturbed environmental state $\tilde{\cal E}
[U_{ab},\sigma_b](\rho_a)$ from the received density matrix ${\cal
E}(\rho_a)$. This requires the existence of a third CPT
transformation ${\cal T}:{\cal D}({\cal H}_a)\rightarrow {\cal
D}({\cal H}_b)$  such that
\begin{eqnarray}
({\cal T}\circ {\cal E})(\rho_a) = \tilde{\cal E}
[U_{ab},\sigma_b](\rho_a)
\;,\label{deg}
\end{eqnarray}
with ``$\circ$'' being the composition of super-operators.
Analogously, generalizing~\cite{SARO},
 we say that $\cal E$ is anti-degradable if,
given an unknown input state
$\rho_a$, a third party (Charlie) which is monitoring the channel
environment can
reconstruct Bob
state ${\cal E}(\rho_a)$ from $\tilde{\cal E}[U_{ab},\sigma_b](\rho_a)$.
As before, this requires the existence
of a CPT map $\overline{\cal T}: {\cal D}({\cal H}_b)
\rightarrow {\cal D}({\cal H}_a)$ which for all inputs gives
\begin{eqnarray}
(\overline{\cal T}\circ \tilde{\cal E}[U_{ab},\sigma_b])(\rho_a) = {\cal E}(\rho_a)\;.
\label{antideg}
\end{eqnarray}
Finally, we say that a map ${\cal E}$ is weakly degradable
(anti-degradable) if it is weakly degradable (anti-degradable)
with respect to some representation~(\ref{unit}).

When the environmental state $\sigma_b$ of Eq.~(\ref{unit}) is
pure [i.e.,  if Eq.~(\ref{unit}) is a Stinespring representation
of $\cal E$] our definition of weak-degradability reduces
 to the degradability property introduced in Ref.~\cite{DEVSHOR}
and implies the additivity of the coherent information of the
channel. In the general case, however, weakly degradable channels
need not to be degradable and Eq.~(\ref{deg}) is only a necessary
condition for degradability. Using the no-cloning
theorem~\cite{NOCLONING}, one can prove that anti-degradable
channels must have null quantum capacity $Q=0$. Indeed, assume by
contradiction $Q>0$. This means that by employing sufficiently
many times the map $\cal E$, Alice will be able to transfer to Bob
a generic unknown state $|\psi\rangle$. However, since the channel
is anti-degradable, everything Bob gets from the channel can also
be reconstructed by Charlie by cascading $\tilde{\cal
E}[U_{ab},\sigma_b]$ with the CPT map $\overline{\cal T}$ of
Eq.~(\ref{antideg}). This implies that at the end of the day both
Bob  and Charlie will have a copy of $|\psi\rangle$, which is
impossible.
\\
Weak-degradability and anti-degradability are not mutually
exclusive properties -- for instance, will see that a beam
splitter channel with transmissivity $1/2$ satisfies both
Eqs.~(\ref{deg}) and (\ref{antideg}). Moreover,  within  the
representation~(\ref{unit}), the weakly complementary of
$\tilde{\cal E}[U_{ab},\sigma_b]$ can be identified with the
original map ${\cal E}$: this implies that ${\cal
E}[U_{ab},\sigma_b]$  is weakly degradable if and only if its
weakly complementary $\tilde{\cal E}[U_{ab},\sigma_b]$ is
anti-degradable.
Finally, an easy to verify but important property
is the fact that maps which are unitarily equivalent to a weakly
degradable channel are also weakly degradable (analogously for
anti-degradability).

\section{One-Mode Gaussian Channels}\label{SEC:ONEMODE}
A CPT transformation $\cal E$ operating on a Bosonic mode
described by the annihilation operator $a$ is said to be Gaussian
if when acting on Gaussian input states $\rho_a$ produces output
Gaussian states~\cite{HW,HOLEVOBOOK}. We recall that the Gaussian
states of the mode $a$ are density matrices $\rho_a$ whose
characteristic function
\begin{eqnarray}
\chi(\mu) = \Tr[ \rho_a \exp(\mu a^\dag -
\mu^* a)] \;\label{Ncharact}
\end{eqnarray}
 is Gaussian, i.e.,
\begin{eqnarray}
\chi(\mu) =\exp[ - \zeta \cdot
\Gamma \cdot \zeta^\dag/2 - \zeta_0 \cdot \zeta^\dag]
\;,
\end{eqnarray}
 with
$\zeta=(\mu^*,-\mu)$, $\Gamma$ being the covariant matrix of
$\rho_a$ and $\zeta_0 $ being the first order momentum of the
distribution~\cite{HOLEVOBOOK,HW,REV1}.

In the following we will focus on one-mode Gaussian channels which
admit a single-mode unitary representation. They can be expressed
as in Eq.~(\ref{unit}) with $\sigma_b$ being a (possibly mixed)
Gaussian state of a {\em single} environmental Bosonic mode
described by the annihilation operator $b$. In this context
$U_{ab}$ describes a linear coupling which performs the
transformation~\cite{HOLEVOBOOK,HW},
\begin{eqnarray}
U_{ab}\; \vec{v} \; U_{ab}^\dag = A
\cdot \vec{v} \label{linear}\;,
\end{eqnarray}
where $\vec{v}^T= (a,a^\dag,b, b^\dag)$ and $A$ being a $4\times
4$ complex symplectic matrix.
In particular, to preserve the commutation relation among
the operators $a$, $a^\dag$, $b$ and $b^\dag$, the matrix
$A$ satisfies the following constraints
\begin{eqnarray}
\sum_{j=1}^4{(-1)^{j+1} |A_{i j}|^2}=1 \label{NEWCONS1}
\end{eqnarray}
for $i=1,3$ and
\begin{eqnarray}
\sum_{j=1}^4{(-1)^{j+1} A_{1 j} A_{3 \ j+(-1)^{j+1}}}&=&0\;,
\nonumber
\\
\sum_{j=1}^4{(-1)^{j+1} A_{1 j} A_{3 j}^{*}}&=&0 \;.
\label{NEWCONS2}
\end{eqnarray}
Almost all the one-mode Gaussian channels can be expressed in this
way. Indeed the only exception to this rule is represented by maps
which are unitarily equivalent to
additive-classical-Gaussian-noise channels~\cite{HOLEVOREP,NUOVO}.
Within the single-mode unitary representation of $\cal E$, the
weakly complementary map~(\ref{unitKconjugate}) of $\cal E$ is
again an one-mode Gaussian channel~\cite{HW,HOLEVONEW} which can
be seen as a transformation which maps ${\cal D}({\cal H}_a)$ into
itself, by introducing an irrelevant isometry which exchanges $a$
and $b$~\cite{SARO}. We will show that the weak-degradability of
the Gaussian map ${\cal E}[U_{ab},\sigma_b]$ with $U_{ab}$ as in
Eq.~(\ref{linear}) depends upon the real  parameter,
\begin{eqnarray}
q \equiv |A_{11}|^2 - |A_{12}|^2
\label{defq}\;,
\end{eqnarray}
with $A_{11}$ and $A_{12}$ being elements of the matrix $A$. The
quantity~(\ref{defq}) is an invariant of the unitary
representation of the map, i.e., it depends on $\cal E$ but not on
the choice of $U_{ab}$ and $\sigma_b$. This property is discussed
in details in Refs.~\cite{HOLEVOREP,NUOVO} (see also
Ref.~\cite{SEW}). Here we will prove that the map ${\cal
E}[U_{ab},\sigma_b]$ is weakly degradable for $q\geqslant 1/2$ and
anti-degradable for $q\leqslant 1/2$ (see Table~\ref{t:table}).

\begin{table}[t]
\begin{tabular}  {|c|c|c|c|}
\hline \hline
Value of $q$ & Equivalent map &    \\ \hline \hline
$q< 0$ & $\tilde{\cal E}[1-q,\sigma_b^\prime]$ & Anti-degradable  \\
&conjugate amplifier & $(Q=0)$   \\
\hline
 $0 < q\leqslant {1}/{2} $ & ${\cal E}[q,\sigma_b^\prime]$
& Anti-degradable  \\
& BS of transmissivity $q$& $(Q=0)$  \\ \hline
$1/2 \leqslant  q < 1$& BS of transmissivity $q$ & \\
 & ${\cal E}[q,\sigma_b^\prime]$ & Weakly degradable
  \\
$1 < q $& amplifier  & (degradable for $\sigma_b^\prime$ pure)\\
\hline
\end{tabular}
\caption{Weak-degradability and anti-degradability conditions for
the one-mode Bosonic channel ${\cal E}[U_{ab},\sigma_b]$. In the
first column we report the value of the characteristic parameter
$q$ of Eq.~(\ref{defq}). In the second column we report the BS or
amplifier map which, according to Eqs.~(\ref{decomposition1}) and
(\ref{decomposition2}), is unitarily equivalent to $\cal E$
($\sigma_b^\prime$ are Gaussian states obtained by properly
squeezing $\sigma_b$). For $q=0$ and $q=1$ the equivalent BS or
amplifier map not always exists~\cite{HOLEVOREP}, still one can
show that these maps are respectively anti-degradable and weakly
degradable. Channels which are anti-degradable have null quantum
capacity. Those which are weakly degradable with $\sigma_b$ pure
(i.e., degradable) have instead additive coherent information. The
case $q=1/2$ is an example of a channel which satisfies both the
weak-degradability condition~(\ref{deg}) and the
anti-degradability condition~(\ref{antideg}). This is a
consequence of the symmetry of
 the fields emerging
from the opposite output ports of a $50/50$ beam-splitter.
\label{t:table}} \end{table}

Without loss of generality, in the following we will assume
$\sigma_b$ to have null first order momentum (it can always be
compensated through a suitable unitary operator acting on the
output of the channel). Another important simplification arises by
considering the one-parameter family of unitaries
$U_{ab}^{(k)}$~(\ref{linear}) associated with beam-splitter (BS)
and amplifier transformations. For $k\in[0,1]$ they are
characterized by the matrix
\begin{eqnarray}
A^{(k)} =\left( \begin{array}{cccc}
\sqrt{k}& 0 & - \sqrt{1-k} & 0  \\
0 & \sqrt{k} & 0 &- \sqrt{1-k} \\
\sqrt{1-k} & 0 &  \sqrt{k} & 0 \\
0 &\sqrt{1-k}& 0& \sqrt{k}
\end{array} \right) , \nonumber \\
\label{MATRIX1}
\end{eqnarray}
which describes
superposition of the modes $a$ and $b$ at the output of a beam-splitter
of transmissivity $k$. For $k\geqslant 1$ instead the $U_{ab}^{(k)}$ are
 characterized by the matrix
\begin{eqnarray}
A^{(k)} =\left( \begin{array}{cccc}
\sqrt{k}& 0& 0&  - \sqrt{k-1} \\
0& \sqrt{k} & - \sqrt{k-1} & 0  \\
0 & - \sqrt{k-1} &\sqrt{k}& 0  \\
- \sqrt{k-1} & 0 & 0 & \sqrt{k}
\end{array} \right), \nonumber \\
\label{MATRIX2}
\end{eqnarray}
which defines an amplification of $a$ with gain parameter $k$.
Notice that in both cases Eq.~(\ref{defq})
yields
$$|A_{11}^{(k)}|^2 - |A_{12}^{(k)}|^2 = k \; .$$
As discussed in Appendix~\ref{appendix}, the corresponding maps
${\cal E}[k,\sigma_b]\equiv {\cal E}[U^{(k)}_{ab},\sigma_b]$ and
their weakly conjugate $\tilde{\cal E}[k,\sigma_b] \equiv
\tilde{\cal E}[U^{(k)}_{ab},\sigma_b]$ can be used to express a
generic one-mode Gaussian channel via proper unitary
transformations, with some remarkable exceptions in the case of
$q=0,1$~\cite{HOLEVOREP}. We can, therefore, prove the
weak-degradability or anti-degradability property of one-mode
Gaussian maps by focusing only on the subset ${\cal
E}[k,\sigma_b]$ (see Sec.~\ref{sec:bsandamp}). Consider, in fact,
a generic Gaussian map of the form ${\cal E}[U_{ab},\sigma_b]$
with the  real parameter $q$ of Eq.~(\ref{defq}) being positive
and $\neq 1$. According to Eq.~(\ref{unit2}) of the Appendix we
can write,
\begin{eqnarray}
{\cal E}[U_{ab},\sigma_b](\rho_a) = S_a \; \big( \; {\cal E}[k=q,
\sigma_b^\prime] (\rho_a) \; \big) \; S_a^\dag \;,
\label{decomposition1}
\end{eqnarray}
with $\sigma_b^\prime \equiv S_b^{\prime} \sigma_b
{S_b^\prime}^{\dag}$ and $S_a$, $S_b^\prime$ being, respectively,
unitary squeezing operators of $a$ and $b$ which depend on $A$ but
not on the input state $\rho_a$. Since squeezed Gaussian states
are Gaussian, the above expression shows that any Gaussian channel
${\cal E}[U_{ab},\sigma_b]$ is unitarily equivalent to an
amplifier channel for $q > 1$ and to a BS channel for $q\in
]0,1[$. As will be shown in the next section, this implies that
${\cal E}[U_{ab},\sigma_b]$ is anti-degradable for $q\in ] 0,1/2]$
and weakly degradable for $q\geqslant 1/2$ and $q\neq 1$. Consider
now the case of maps with $q$ of Eq.~(\ref{defq}) being negative.
Here Eq.~(\ref{decomposition1}) is replaced by
\begin{eqnarray}
{\cal E}[U_{ab},\sigma_b](\rho_a) = S_a \; \big( \; \tilde{\cal E}[1- q, \sigma_b^\prime]
(\rho_a) \; \big) \; S_a^\dag \;,
\label{decomposition2}
\end{eqnarray}
where, again,  $S_a$ and $\sigma_b^\prime$ are, respectively, a
squeezing operator and a Gaussian state [in writing
Eq.~(\ref{decomposition2}) an isometry $a \leftrightarrow b$ is
implicitly assumed]. Since $1-q > 1$, Eq.~(\ref{decomposition2})
shows that $\cal E$ is unitarily equivalent to the weakly
conjugate map of the amplifier channel ${\cal E}[1-q,
\sigma_b^\prime]$. As discussed in the following this is
equivalent to say that ${\cal E}[U_{ab},\sigma_b]$ with negative
$q$ are always anti-degradable. Finally, for $q=0$ and $q=1$ the
channel is, respectively, anti-degradable and weakly degradable.
The analysis of these maps is slightly more complex since it is
not always possible to describe them in terms of BS/amplifier
channels~\cite{HOLEVOREP} (details are given in
Ref.~\cite{NUOVO}).

\section{BS and Amplifier maps}\label{sec:bsandamp}

In this section we analyze the weak-degradability properties of
the BS and amplifier maps.

According to the definition of the matrix $A^{(k)}$ it follows
that the map ${\cal E}[k,\sigma_b]$ operates on a generic (not
necessarily Gaussian) state $\rho_a$  by transforming its
characteristic function $\chi(\mu)$ as follows:
\begin{eqnarray}
\chi(\mu) \rightarrow \chi^\prime(\mu) = \left\{
\begin{array}{lll}
\chi(\sqrt{k} \mu) \; \xi( \sqrt{1-k}\mu) & &k\in [0,1]\\\\
\chi(\sqrt{k} \mu) \; \xi( - \sqrt{k-1} \mu^*) & &k \geqslant 1\;,
\end{array}
\right. \label{chi1}
\end{eqnarray}
with
\begin{eqnarray}
\xi(\mu) = \Tr [ \sigma_b \exp( \mu b^\dag - \mu^* b)]
\label{NEWxi} \;,
\end{eqnarray}
 being the Gaussian characteristic function of the
environment state $\sigma_b$ which, as previously discussed, is
assumed to have null first order momentum. Analogously the weakly
complementary map
 $\tilde{\cal E}[k,\sigma_b]$
produces the transformation,
\begin{eqnarray}
\chi(\mu) \rightarrow \chi^\prime(\mu) = \left\{
\begin{array}{lll}
\chi(- \sqrt{1-k}\mu) \; \xi( \sqrt{k}\mu) & & k\in [0,1]\\\\
\chi( - \sqrt{k-1} \mu^*) \; \xi(\sqrt{k} \mu) & &k \geqslant 1\;.
\end{array}
\right. \label{chi2}
\end{eqnarray}
We now show that ${\cal E}[k,\sigma_b]$ is weakly degradable for
$k\geqslant 1/2$ and anti-degradable for $k\leqslant 1/2$.

Consider first the amplifier case where $k\geqslant 1$. To show
that ${\cal E}[k,\sigma_b]$ satisfies the weak-degradability
condition~(\ref{deg}) we define the quantity $k^\prime \equiv  (2
k -1 )/k$ and notice that this is always greater than or equal to
$1$. Our claim is that one can identify the map ${\cal T}$ of
Eq.~(\ref{deg}) with the weakly complementary
map~(\ref{unitKconjugate})
 of an Amplifier of gain
$k^\prime$, i.e., ${\cal T} ={\tilde {\cal E}} [
k^\prime,\sigma_b]$. This can be verified by studying how ${\tilde
{\cal E}} [ k^\prime,\sigma_b] \circ {\cal E}[k, \sigma_b]$ acts
on a generic state $\rho_a$. Combining Eqs.~(\ref{chi1}),
(\ref{chi2}) it follows that the characteristic function
$\chi(\mu)$ of $\rho_a$ is transformed into
\begin{eqnarray}
&&\chi( - \sqrt{k (k^\prime-1)} \mu^*) \; \xi(\sqrt{(k^\prime -1)(k-1)} \mu) \; \xi(\sqrt{k^\prime} \mu) \nonumber \\
&&\quad \quad  =  \chi( - \sqrt{k (k^\prime-1)} \mu^*)\; \xi( \sqrt{(k^\prime-1)(k-1)+k^\prime}\mu )\nonumber \\
&&\quad \quad =  \chi( - \sqrt{k-1} \mu^*) \;\xi( \sqrt{k}\mu )
\label{cal1}\;,
\end{eqnarray}
where we used the properties of the Gaussian function $\xi$  and
the identity $k(k^\prime-1) = k-1$. By comparison with
Eq.~(\ref{chi2}), we notice that ${\tilde {\cal E}} [
k^\prime,\sigma_b] \circ {\cal E}[k, \sigma_b]$ operates on
$\rho_a$ as ${\tilde {\cal E}}[k, \sigma_b]$. Since this is true
for all $\rho_a$ we get
\begin{eqnarray}
{\tilde {\cal E}}[k, \sigma_b] = {\tilde {\cal E}} [ k^\prime,\sigma_b] \circ
{\cal E}[k, \sigma_b] \label{composition3}\;,
\end{eqnarray}
proving the thesis.

Consider now the BS case where $k\in [0,1]$. Here we distinguish
two different regimes. For $k\in [1/2,1]$ the channel ${\cal
E}[k,\sigma_b]$  is still weakly degradable and satisfies
Eq.~(\ref{composition3}), the only difference being that now
${\tilde {\cal E}}[k^\prime,\sigma_b]$ represents the weakly
complementary of a beam-splitter map of transmissivity $k^\prime =
(2k-1)/k \in [0,1]$. The formal proof goes as in Eq.~(\ref{cal1}),
which now becomes

\begin{eqnarray}
&& \chi( - \sqrt{k (1-k^\prime)} \mu) \;
\xi(-\sqrt{(1-k^\prime)(1-k)} \mu) \; \xi(\sqrt{k^\prime} \mu)\nonumber \\
&&\quad = \chi( - \sqrt{k (1-k^\prime)} \mu)\; \xi(
\sqrt{(1-k^\prime)(1-k)+k^\prime}\mu ) \nonumber \\ &&\quad =
\chi( - \sqrt{1-k} \mu) \;\xi( \sqrt{k}\mu ) \label{cal2}\;.
\end{eqnarray}

For $k\in[0,1/2]$ instead we can show that ${\cal E}[k,\sigma_b]$
is anti-degradable by observing that it satisfies the
condition~(\ref{antideg}) with $\overline{\cal T}$ being the
weakly complementary $\tilde{\cal E}[k^{\prime\prime},\sigma_b]$
of a BS channel of transmissivity $k^{\prime\prime} =
(1-2k)/(1-k)\in[0,1]$, i.e.,
\begin{eqnarray}
{\cal E}[k, \sigma_b] = {\tilde {\cal
E}} [ k^{\prime\prime},\sigma_b] \circ \tilde{\cal E}[k, \sigma_b]\;.
\label{NUOVA33}
\end{eqnarray}
The proof is again obtained through Eqs.~(\ref{chi1}) and
(\ref{chi2}) by showing that the transformations on a generic
$\chi(\mu)$ induced
 by ${\tilde {\cal E}} [ k^{\prime\prime},\sigma_b] \circ
\tilde{\cal E}[k, \sigma_b]$ and by
${\cal E}[k, \sigma_b]$ coincide.

Indeed, one has
\begin{eqnarray}
&& \chi( \sqrt{(1-k) (1-k^{''})} \mu) \; \xi(\sqrt{(1-k^{''})k}
\mu) \; \xi(\sqrt{k^{''}} \mu) \nonumber \\
&&\quad = \chi( \sqrt{(1-k) (1-k^{''})} \mu)
\; \xi( \sqrt{(1-k^{''})k+k^{''}}\mu ) \nonumber \\
&&\quad = \chi( \sqrt{k} \mu) \;\xi( \sqrt{1-k}\mu ).
\end{eqnarray}

\section{Conclusions}
A new property of quantum channels (i.e., weak-degradability) is
introduced by exploiting a more ``physical'' picture of the noise
evolution (i.e., interaction with a thermal-like environment). We
prove that with the exception of the
additive-classical-Gaussian-noise channels~\cite{HOLEVOREP}, all
one-mode Gaussian maps are weakly degradable or anti-degradable.
In the latter case this implies that their  quantum capacity $Q$
must be null. In the former case instead this yields the
additivity of their coherent information
 under the condition that the
representation~(\ref{unit}) upon which weak-degradability was
derived, possesses single-mode pure environmental state. For the
sake of completeness we mention that after the submission of our
manuscript, Wolf {\em et al.} posted a couple of interesting
papers \cite{WOLF} where, adopting an approach similar to ours,
the quantum capacities of several quantum channels were explicitly
solved (including the one-mode Gaussian channels which here we
show to have additive coherent information).

\acknowledgments We thank Professor A. S. Holevo for pointing out
a flaw in our original argument. This led us to the definition of
weak-degradability and helped  us in clarifying the possibility of
describing {\em almost} all one-mode Gaussian maps in terms of
single-mode environment representations. We also thank Professor
R. Fazio for comments. This work was supported by the Quantum
Information research program of Centro di Ricerca Matematica Ennio
De Giorgi of the Scuola Normale Superiore.

\appendix

\section{Decomposition rules}~\label{appendix}
Here we give an explicit derivation of the decomposition
rules~(\ref{decomposition1}) and (\ref{decomposition2}) which
allow us to express any generic one-mode Gaussian map with $q\neq
0,1$ in terms of BS or amplifier channels. For the sake of clarity
we will analyze separately the cases $q\in ]0,1[$, $q>1$ and
$q<0$.

\subsection{Maps with $q\in]0,1[$}

Consider first the case of one-mode Gaussian channel of the form
${\cal E}[U_{ab},\sigma_b]$ with the  real parameter $q$ of
Eq.~(\ref{defq}) being positive and smaller than $1$. Under this
condition, apart from redefining  the phasis of $a$ and $b$, the
elements $A_{1j}$ of the matrix~(\ref{linear}) can be
parameterized as follows
\begin{eqnarray}
A_{11} &=& \sqrt{q}\; \cosh r \;, \nonumber \\
A_{12} &=& \sqrt{q} \; e^{i\varphi} \sinh r \;, \nonumber\\
A_{13} &=& - \sqrt{1-q}\; \cosh s \;,  \nonumber \\
A_{14} &=& - \sqrt{1-q} \; e^{i\psi} \sinh s \;, \label{NPRIME}
\end{eqnarray}
where $r,s,\varphi$, and $\psi$ are real quantities and where the
last two expressions come  from the constraint~(\ref{NEWCONS1}).
Let us then introduce the (unitary) squeezing
transformations~\cite{WMILBURN}:
\begin{eqnarray}
S_a(r;\varphi)\; a \;S_a^\dag(r;\varphi) &=& a \; \cosh r +
e^{i\varphi} \; a^\dag \;\sinh r \;, \nonumber
\\
S_b(s;\psi) \; b \; S_b^\dag(s;\psi) &=& b\; \cosh s + e^{i\psi}\;
b^\dag\; \sinh s \;. \label{squeezing}
\end{eqnarray}
On one hand, they allow us to write
\begin{eqnarray}
(S_a^\dag \otimes S_b^\dag) \; a^\prime \;
(S_a \otimes S_b) &=& \sqrt{q} \; a -\sqrt{1-q} \; b \nonumber \\
&=& U_{ab}^{(q)} \; a \; \left[U_{ab}^{(q)}\right]^\dag  \;,
\label{NEWUNO}
\end{eqnarray}
where $U_{ab}^{(q)}$ is the BS transformation defined as in
Eq.~(\ref{MATRIX1}) while $a^\prime=U_{ab} \; a \; U_{ab}^\dag$
represents the evolution of $a$ under the unitary $U_{ab}$ of
${\cal E}[U_{ab},\sigma_b]$. On the other hand, we get
\begin{eqnarray}
(S_a^\dag \otimes S_b^\dag) \; b^\prime \;
(S_a \otimes S_b) &=& \overline{A}_{21} a +
\overline{A}_{22} a^\dag +\overline{A}_{23} b +
 \overline{A}_{24} b^\dag1\;, \nonumber \\
\label{NEWEQ33}
\end{eqnarray}
with $b^\prime = U_{ab}\; b \;U_{ab}^\dag$ and with
$\overline{A}_{2j}$
 being complex parameters which
satisfies the symplectic conditions analogous to those of
Eqs.~(\ref{NEWCONS1}) and (\ref{NEWCONS2}), i.e.,
\begin{eqnarray}
|\overline{A}_{21}|^2 - |\overline{A}_{22}|^2
+|\overline{A}_{23}|^2 - |\overline{A}_{24}|^2 &=&1\;, \nonumber \\
\sqrt{q} \; \overline{A}_{21} - \sqrt{1-q} \;\overline{A}_{23} &=&
0\;, \nonumber
\\
\sqrt{q} \; \overline{A}_{22} - \sqrt{1-q} \; \overline{A}_{24}
&=& 0 \;.
\end{eqnarray}
Equation~(\ref{NEWEQ33}) can be cast in a more compact form by
properly parameterizing the $\overline{A}_{2j}$;
\begin{eqnarray}
\overline{A}_{21} &=& \sqrt{1-q} \; \cosh (t) \; e^{i \phi}\;, \nonumber \\
\overline{A}_{22} &=& \sqrt{1-q} \; \sinh (t) \; e^{i \phi^\prime}\;, \nonumber \\
\overline{A}_{23} &=& \sqrt{q} \; \cosh (t) \; e^{i \phi}\;, \nonumber \\
\overline{A}_{24} &=& \sqrt{q} \; \sinh (t) \; e^{i \phi^\prime}
\;, \label{PARA}
\end{eqnarray}
with $t, \phi$, and $\phi^\prime$ real. This yields
\begin{eqnarray}
(S_a^\dag \otimes S_b^\dag) \; b^\prime \; (S_a \otimes S_b) =
 e^{i\phi} \; U_{ab}^{(q)}
\left( \; S_b^\prime \; b\; {S_b^\prime}^\dag\; \right)
\left[U_{ab}^{(q)}\right]^\dag \;, \nonumber
\\  \label{NEWUNO1}
\end{eqnarray}
where $S_b^\prime\equiv S_b(t,\phi^\prime-\phi)$ is a squeezing
operator~(\ref{NEWEQ33}) acting on $b$ and where $U_{ab}^{(q)}$ is
the BS unitary coupling of Eq.~(\ref{NEWUNO}). By absorbing the
phase $\phi$ into the definition of $b^\prime$ and by noticing
that $S_b^\prime$ does not affect $a$, Eqs.~(\ref{NEWUNO}),
(\ref{NEWUNO1}), and (\ref{linear}) give
\begin{eqnarray}
U_{ab}\; \vec{v} \; U_{ab}^\dag  = (S_a \otimes S_b) \;
U_{ab}^{(q)} S_b^\prime \; \vec{v} \; {S_b^\prime}^\dag  \;
[U_{ab}^{(q)}]^\dag \;
(S_a \otimes S_b)^\dag \;,\nonumber \\
\label{NEW100}
\end{eqnarray}
which allows us to decompose $U_{ab}$ as the following product:
\begin{eqnarray}
U_{ab} = (S_a \otimes S_b) \; U_{ab}^{(q)} S_b^\prime
\label{NEW1001}\;.
\end{eqnarray}
Replacing this into Eq.~(\ref{unit}) we finally get
\begin{eqnarray}
{\cal E}[U_{ab},\sigma_b](\rho_a) &=& S_a \Tr_b[ S_b U_{ab}^{(q)}
(\rho_a\otimes \sigma_b^\prime) [U_{ab}^{(q)}]^\dag S_b^\dag]
S_a^\dag
\nonumber \\
&=& S_a \Tr_b[ U_{ab}^{(q)} (\rho_a\otimes \sigma_b^\prime)
[U_{ab}^{(q)}]^\dag ]
S_a^\dag \nonumber \\
&=&  S_a \; \big( \; {\cal E}[k=q, \sigma_b^\prime] (\rho_a) \;
\big) \; S_a^\dag \label{unit2} \;.
\end{eqnarray}
In this expression the $S_a$ was brought out of the trace since it
is acting on $a$. Vice versa, $S_b$ has been simplified by
exploiting the invariance of the trace under unitary
transformation. Finally, the Gaussian state $\sigma_b^\prime$
 is the squeezed version under $S_b^\prime$ of the environmental
state $\sigma_b$, i.e.,
\begin{eqnarray}
\sigma_b^\prime \equiv S_b^\prime \sigma_b {S_b^\prime}^\dag \;.
\end{eqnarray}
Equation~(\ref{unit2}) shows that, for $q\in]0,1[$ the map ${\cal
E}[U_{ab},\sigma_b]$ is unitary equivalent to the BS channel
${\cal E}[k=q, \sigma_b^\prime]$.

\subsection{Maps with $q>1$}

For $q$ greater than one Eqs.~(\ref{NEW1001}) and (\ref{unit2})
still hold: the only difference being that now $U_{ab}^{(q)}$
represents an amplifier map defined by the matrix of
Eq.~(\ref{MATRIX2}). This can be shown following the same
derivation of the case $q\in]0,1[$ by replacing the
parameterizations~(\ref{NPRIME}) and (\ref{PARA}) with
\begin{eqnarray}
A_{11} &=& \sqrt{q}\; \cosh r \;, \nonumber \\
A_{12} &=& \sqrt{q} \; e^{i\varphi} \sinh r \;, \nonumber\\
A_{13} &=& - \sqrt{q-1}\; e^{- i\psi} \sinh s \;,  \nonumber \\
A_{14} &=& - \sqrt{q-1} \;  \cosh s \;, \label{NPRIME1}
\end{eqnarray}
and
\begin{eqnarray}
\overline{A}_{21} &=& - \sqrt{q-1} \; \sinh (t) \; e^{i \phi^\prime}\;, \nonumber \\
\overline{A}_{22} &=& - \sqrt{q-1} \; \cosh (t) \; e^{i \phi}\;, \nonumber \\
\overline{A}_{23} &=& \sqrt{q} \; \cosh (t) \; e^{i \phi}\;, \nonumber \\
\overline{A}_{24} &=& \sqrt{q} \; \sinh (t) \; e^{i \phi^\prime}
\;. \label{PARA2}
\end{eqnarray}

\subsection{Maps with $q<0$}
To analyze the channels ${\cal E}[U_{ab},\sigma_b]$
 with  $q$ negative it is useful to
introduce a isometry $\Xi_{ab}=\Xi_{ab}^\dag$ which transforms $a$
in $b$ and vice versa while leaving the vacuum state invariant,
i.e., $\Xi_{ab} \;a\; \Xi_{ab} = b$, $\Xi_{ab}\;  b \; \Xi_{ab} =a
$, and $\Xi_{ab} |\O\rangle = |\O\rangle$. This  is a unitary
transformation which for  any bounded operator $\Theta_{ab}$ on
${\cal H}_a\otimes {\cal H}_b$ satisfies the identity
\begin{eqnarray}
 \Tr_b [ \Xi_{ab} \Theta_{ab} \Xi_{ab} ]\otimes \openone_b  = \Xi_{ab}
\left( \openone_a \otimes \Tr_a [  \Theta_{ab} ] \right) \Xi_{ab}.
\label{TRACCIA}
\end{eqnarray}
Consider, then, the unitary transformation $\Xi_{ab} U_{ab}$ with
$U_{ab}$ being the unitary coupling associated with ${\cal
E}[U_{ab},\sigma_b]$. From Eq.~(\ref{linear}) it follows

\begin{eqnarray}
( \Xi_{ab}\; U_{ab})  \; \vec{v} \;( U_{ab}^\dag \; \Xi_{ab} ) =
\tilde{A}\cdot \vec{v} \label{notice}
\end{eqnarray}

where $\tilde{A}$ is a $4\times4$ matrix which is obtained by
shifting by $2$ the columns of the matrix $A$ which describes the
unitary $U_{ab}$, i.e., $\tilde{A}_{ij} = A_{i , j\oplus 2}$ where
$\oplus$ represents the sum modulus $4$. From the
constraint~(\ref{NEWCONS1}) it then follows that the
coefficient~(\ref{defq}) of $\tilde{A}$ is greater than $1$, i.e.,

\begin{eqnarray}
\tilde{q} &=& |\tilde{A}_{11}|^2 - |\tilde{A}_{12}|^2
= |{A}_{13}|^2 - |{A}_{14}|^2 \nonumber \\
&=& 1 - (|{A}_{11}|^2 - |{A}_{12}|^2) = 1-q >1 \;.\label{EQUO}
\end{eqnarray}

We can then use the previous section to show that there exist
squeezing transformations $S_a$, $S_b$, and $S_b^\prime$ which
allows us to write $\Xi_{ab} \; U_{ab} = (S_a \otimes S_b) \;
U_{ab}^{(\tilde{q})} S_b^\prime$ with $U_{ab}^{(\tilde{q})}$ being
an Amplifier coupling~(\ref{MATRIX2}). Therefore, we get
\begin{eqnarray}
U_{ab} = \Xi_{ab} \; (S_a \otimes S_b) \; U_{ab}^{(\tilde{q})}
S_b^\prime \label{NEW10011}\;.
\end{eqnarray}
Exploiting the identity~(\ref{TRACCIA}) this yields
\begin{widetext}
\begin{eqnarray}
{\cal E}[U_{ab},\sigma_b] (\rho_a)  \otimes \openone_b &=&  \Tr_b
[ \Xi_{ab} \; (S_a \otimes S_b) \; U_{ab}^{(\tilde{q})} S_b^\prime
(\rho_a\otimes \sigma_b) {S_b^\prime}^\dag
\left[U_{ab}^{(\tilde{q})}\right]^\dag
 (S_a \otimes S_b)^\dag  \; \Xi_{ab} ]
 \otimes \openone_b \nonumber \\
&=& \Xi_{ab} ( \openone_a \otimes
  \Tr_a [  (S_a \otimes S_b)
 U_{ab}^{(\tilde{q})}
\times (\rho_a\otimes \sigma_b^\prime)
\left[U_{ab}^{(\tilde{q})}\right]^\dag
 (S_a  \otimes S_b )^\dag ]  ) \Xi_{ab} \nonumber \\
&=& \Xi_{ab} ( \openone_a \otimes
 S_b  \Tr_a [
 U_{ab}^{(\tilde{q})} (\rho_a\otimes \sigma_b^\prime) \left[U_{ab}^{(\tilde{q})}\right]^\dag
 ] S_b^\dag   ) \Xi_{ab} \nonumber
= \Xi_{ab} ( \openone_a \otimes
 S_b \; \tilde{\cal E}[1-q ,\sigma_b^\prime]
 (\rho_a) \;  S_b^\dag   ) \Xi_{ab}\;,
\end{eqnarray}
\end{widetext}
where we used the fact that $\Tr_a [
 U_{ab}^{(\tilde{q})} (\rho_a\otimes \sigma_b^\prime)
\left[U_{ab}^{(\tilde{q})}\right]^\dag
 ]$ is the weakly complementary channel $\tilde{\cal E}[\tilde{q} ,\sigma_b^\prime]$
of an amplifier with coupling  $U_{ab}^{(\tilde{q})}$ and the
 identity $\tilde{q} = 1- q$.
Finally, the above expression can be cast in the less formal but
certainly simpler form~(\ref{decomposition2}) where  the isometry
$\Xi_{ab}$ is implicitly assumed.

\end{document}